\documentclass{aastex63}
\usepackage{CJKutf8}
\usepackage[caption=false]{subfig}
\usepackage{graphicx}
\usepackage{multirow}
\usepackage{booktabs}
\usepackage{amsmath,amssymb}
\usepackage[T1]{fontenc}

\newcommand{\Ni}{{$^{56}$Ni }}

\begin{document}

\title{The UV Excesses of Supernovae and the Implications for Studying Supernovae and Other Optical Transients}

\correspondingauthor{Shan-Qin Wang \begin{CJK*}{UTF8}{gbsn}(王善钦)\end{CJK*}}
\email{shanqinwang@gxu.edu.cn}

\author{Tao Wang \begin{CJK*}{UTF8}{gbsn}(王涛)\end{CJK*}}
\affiliation{Guangxi Key Laboratory for Relativistic Astrophysics,
School of Physical Science and Technology, Guangxi University, Nanning 530004,
China}

\author{Shan-Qin Wang \begin{CJK*}{UTF8}{gbsn}(王善钦)\end{CJK*}}
\affiliation{Guangxi Key Laboratory for Relativistic Astrophysics,
School of Physical Science and Technology, Guangxi University, Nanning 530004,
China}

\author{Wen-Pei Gan \begin{CJK*}{UTF8}{gbsn}(甘文沛)\end{CJK*}}
\affiliation{Guangxi Key Laboratory for Relativistic Astrophysics,
School of Physical Science and Technology, Guangxi University, Nanning 530004,
China}
\affiliation{Nanjing Institute of Astronomical Optics \& Technology, Nanjing 210042, China}

\begin{abstract}

Supernovae (SNe), kilonovae (KNe), tidal disruption events (TDEs), optical afterglows
of gamma ray bursts (GRBs), and many other optical transients are important phenomena
in time-domain astronomy. Fitting the multi-band light curves (LCs) or the synthesized
(pseudo-)bolometric LCs can be used to constrain the physical properties of
optical transients. The (UV absorbed) blackbody module is one of
the most important modules used to fit the multi-band LCs of optical
transients having (UV absorbed) blackbody spectral energy distributions (SEDs).
We find, however, that the SEDs of some SNe show UV excesses, which cannot be fitted
by the model including a (UV absorbed) blackbody module. We construct the bolometric
LCs and employ the (cooling plus) \Ni model to fit the constructed bolometric
LCs, obtaining decent fits. Our results demonstrate
that the optical transients showing UV excesses cannot
be fitted by the multi-band models that include (UV-absorbed) blackbody module,
but can be well modeled by constructing and fitting their bolometric LCs.

\end{abstract}


\section{Introduction}
\label{sec:intro}

In the past two decades, the wide-field optical survey telescopes discovered
supernovae (SNe), kilonovae (KNe), tidal disruption events (TDEs), optical
afterglows of Gamma-ray bursts (GRBs), and other optical transients.
To constrain the physical properties of the optical transients,
the method of fitting the multi-band light curves (LCs) is widely adopted
to model the photometry of superluminous SNe \citep{Nich2017b,Moriya2018},
stripped-envelope SNe \citep{Zheng2022}, TDEs \citep{Mockler2019}, KNe \citep{Villar2017},
and (super)luminous rapidly evolving optical transients or SNe \citep{Wang2019,Wang+Gan2022}.

The (UV absorbed) blackbody module
(e.g., \citealt{Chomiuk2011,Nich2017a,Nich2017b,Prajs2017})
is one of the most important modules used to fit the multi-band LCs of optical
transients, since the spectral energy distributions (SEDs) of the
optical transients listed above (except for the optical afterglows of GRBs)
can be approximately described by the (UV absorbed) blackbody function.

In this paper, we demonstrate that the SEDs of five SNe
(SN~2010jr, SN~2012au, SN~2013ak, SN~2013df, and SN~2014ad) show
UV excesses which cannot be described by UV absorbed blackbody
or standard blackbody function. The UV-optical LCs cannot
be fitted by the multi-band model, while the optical photometry can
be well fitted by the same model.

To better constrain the physical properties, we construct and model
the bolometric LCs of the five SNe, getting good fits.
In Section \ref{UV_excesses}, we demonstrate that the UV-optical SEDs of
the five SNe show UV excesses by fitting the SEDs and multi-band LCs.
In Section \ref{bolometric_LCs}, we construct and model the bolometric LCs of the five SNe.
We discuss our results and draw some conclusions in Section \ref{sec:discussion}.
Throughout the paper, we assume $\Omega_m = 0.315$, $\Omega_\Lambda = 0.685$,
and H$_0 = 67.3$\,km\,s$^{-1}$\,Mpc$^{-1}$ \citep{Planck2014}. The values of the Milky Way
reddening ($E_\mathrm{B-V}$) of all events are from \cite{Schlafly2011}.

\section{The SED and LC Fits}
\label{UV_excesses}

The photometric data and the detailed information (e.g., the positions, SN types, the values of their redshifts, and so on)
of the five SNe, SN~2010jr, SN~2012au, SN~2013ak, SN~2013df, and SN~2014ad, are from \cite{Brown2014}, \cite{Milisavljevic2013}, \cite{Brown2014},
\cite{Morales-Garoffolo2014,Brown2014}, and \cite{Sahu2018} via the collection of {\it Open Supernova Catalog} \citep{Guillochon2017},
respectively.

\begin{figure}
	\centering
	\includegraphics[width=0.30\textwidth,angle=0]{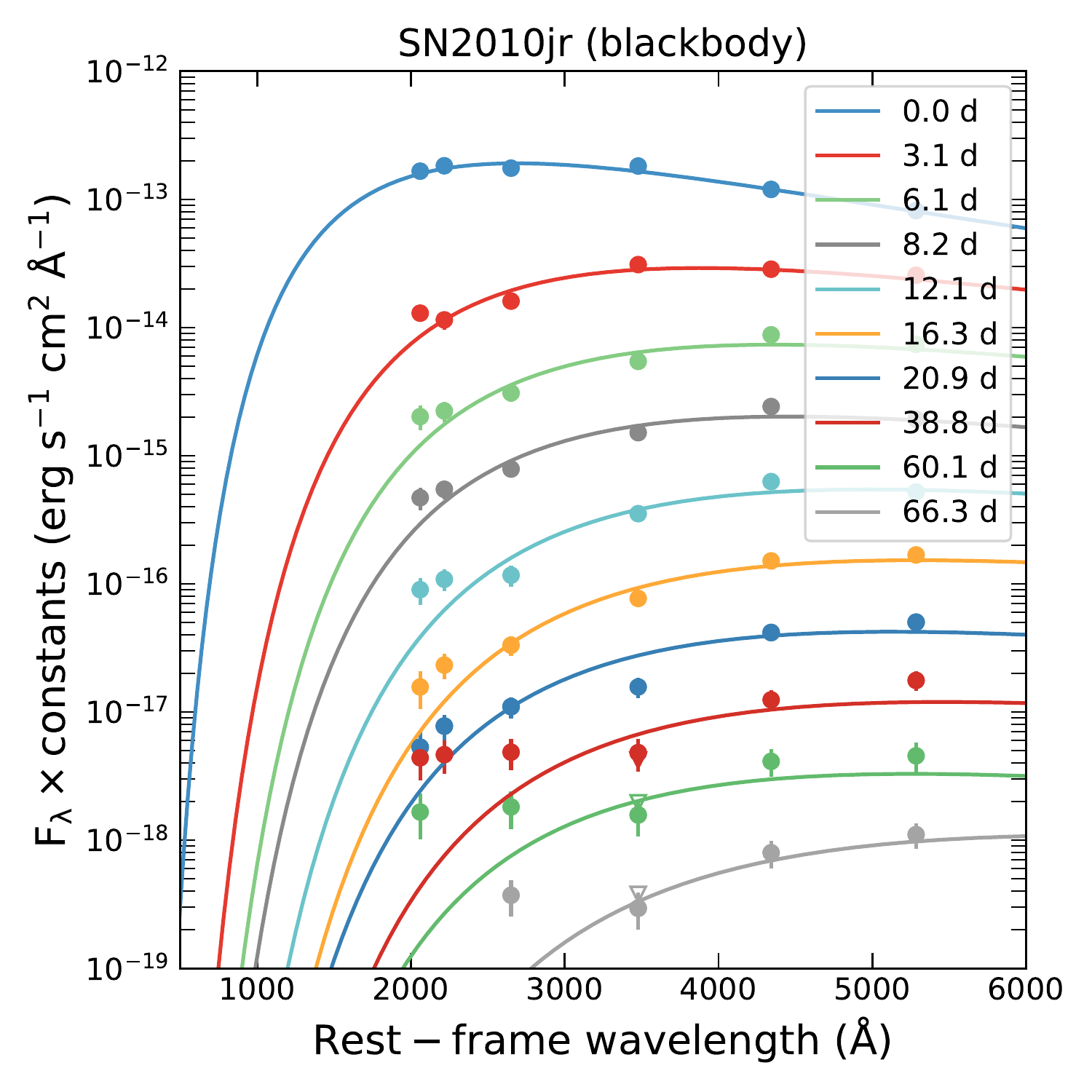}
	\includegraphics[width=0.30\textwidth,angle=0]{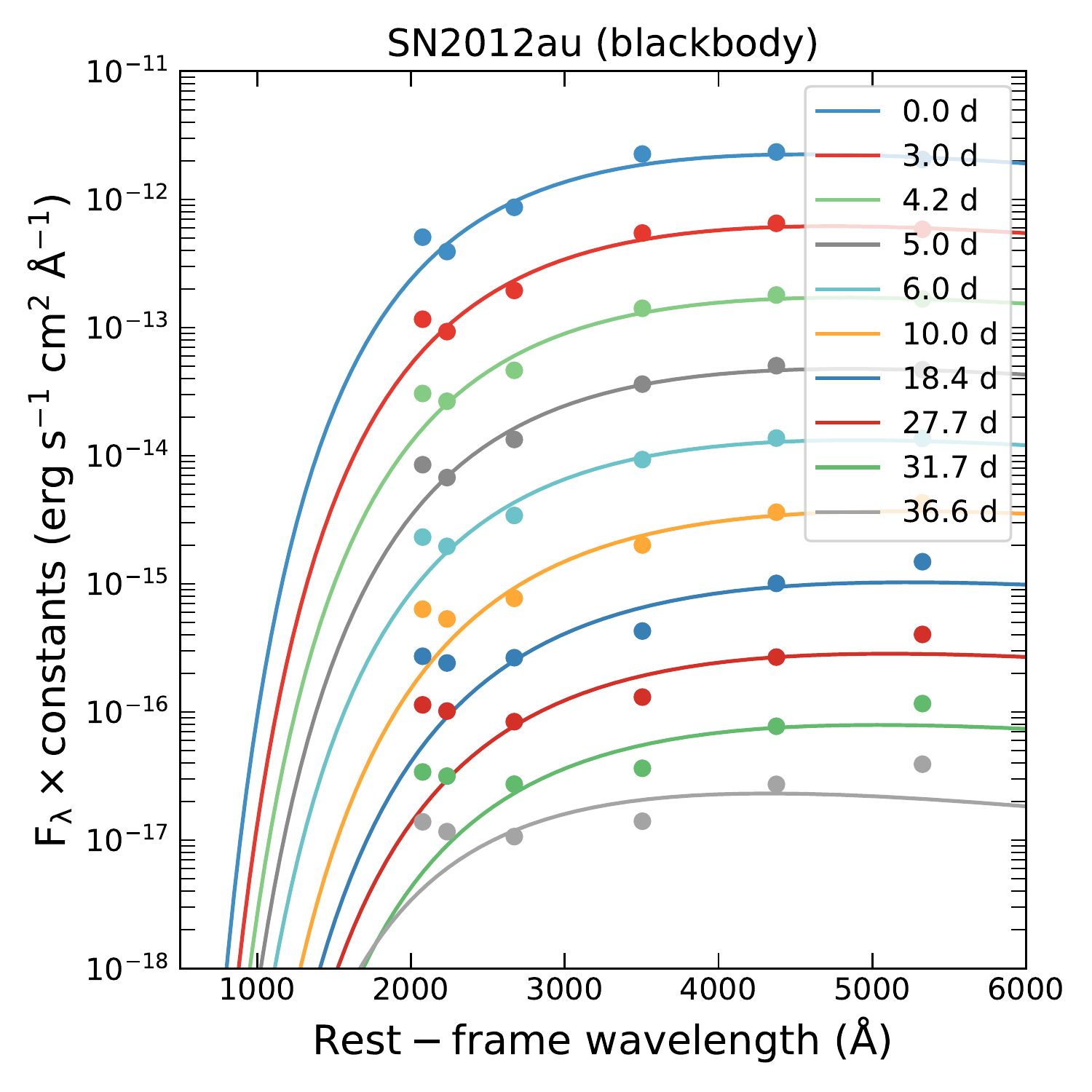}	
	\includegraphics[width=0.30\textwidth,angle=0]{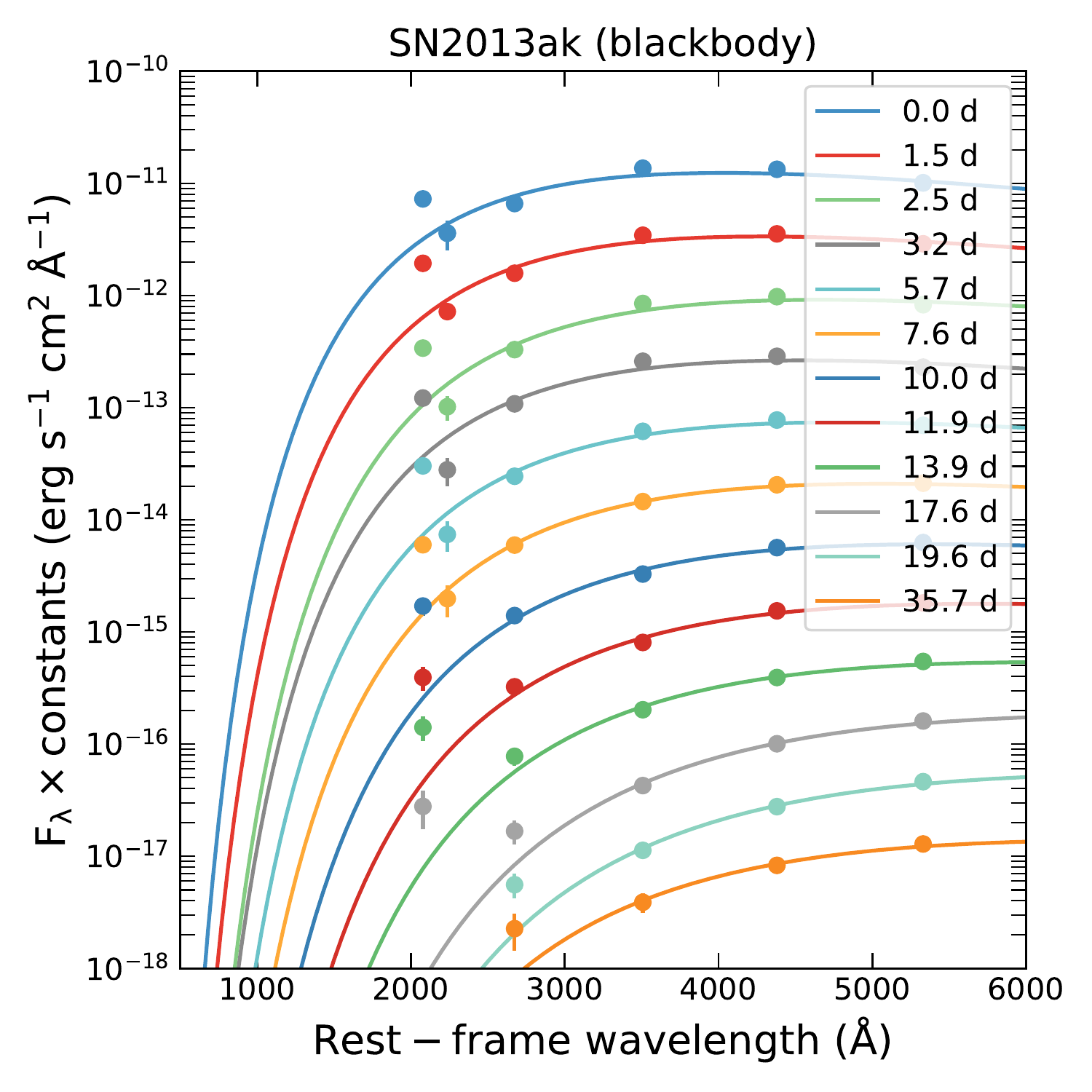}
	\includegraphics[width=0.30\textwidth,angle=0]{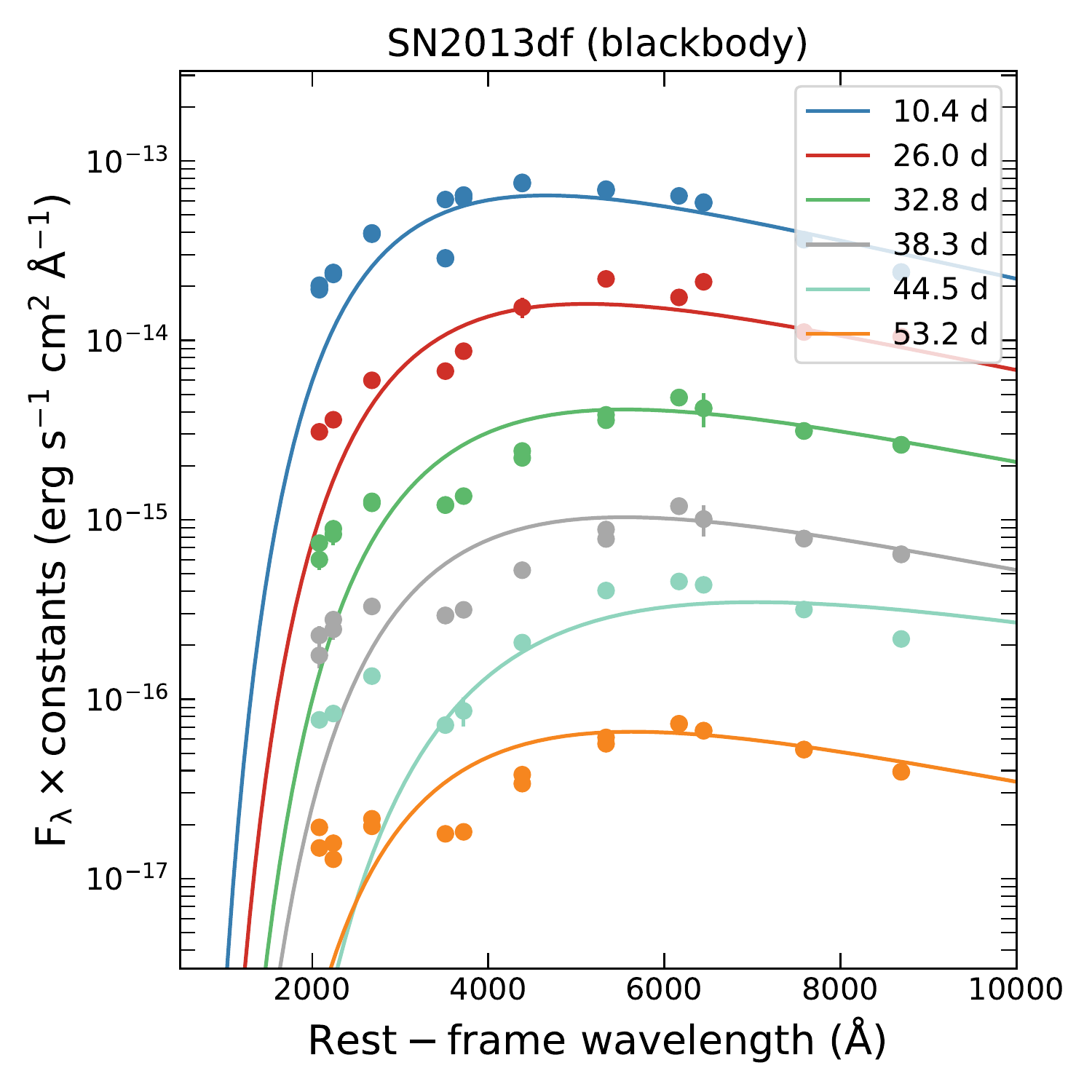}
	\includegraphics[width=0.30\textwidth,angle=0]{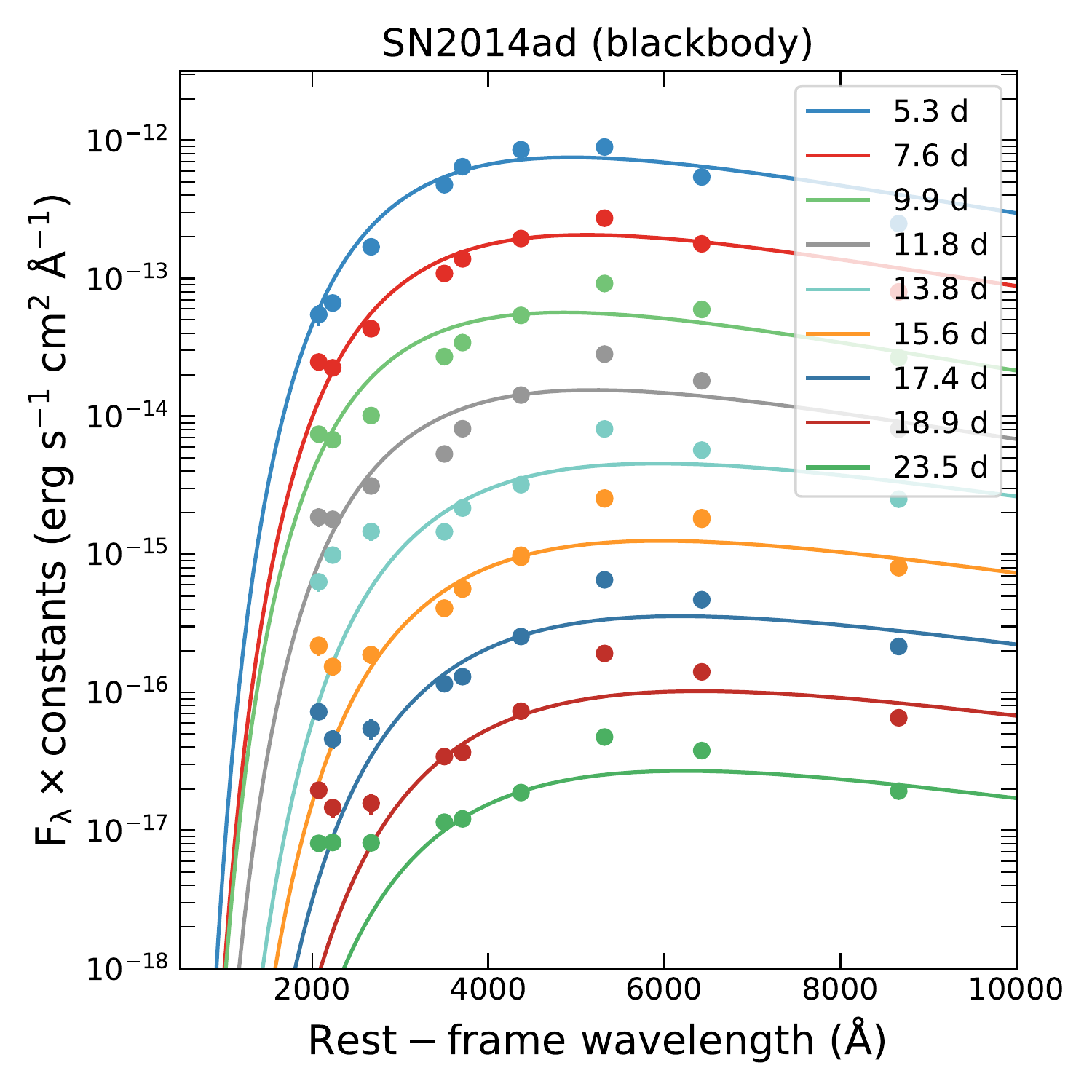}
	\caption{The best fits of the SEDs of the SNe at different epochs.}
\label{Fig:SED}
\end{figure}

First, we use the standard blackbody function
($F_{\nu}=(2\pi h\nu^3/c^2)(e^{{h\nu}/{k_bT_{\rm ph}}}-1)^{-1}\frac{R_{\rm ph}^2}{D^2_{\rm L}}$,
$T_{\rm ph}$ is the temperature of the SN photosphere, $R_{\rm ph}$ is the
radius of the SN photosphere, $D_{\rm L}$ is the luminosity distance of the SN)
to fit their UV-optical SEDs at all epochs, see Figure \ref{Fig:SED}.
\footnote{The linear interpolation is performed in some epochs to obtain the SEDs at the same epochs.
We don't include some optical photometric data of SN~2013df at 150-250 d, since they cannot be used
to construct SEDs.}

We find that the SEDs at all epochs (for SN~2012au, SN~2013ak, and SN~2013df) or
the late epochs (for SN~2010jr and SN~2014ad) of the SNe show UV excesses relative to the standard blackbody
model. The UV excesses of the SEDs indicate that the UV-optical photometry
cannot be fitted by the multi-band model that includes the
standard or UV absorbed blackbody assumption.

To support the statements above, we adopt the \Ni or the cooling
plus \Ni model to fit the LCs of the three single-peaked SNe (SN~2012au, SN~2013ak, and SN~2014ad)
and the two double-peaked SNe (SN~2010jr and SN~2013df), respectively.
\footnote{Since they are stripped-envelope SNe (IIb, Ib and Ic) with normal peak luminosities,
the energy from the recombination of the ionized hydrogen or the magnetar spinning-down can be neglected.}
The details of the \Ni model can be found in the references of \cite{Wang+Gan2022}, while
the details of the cooling model can be found in \cite{Piro2021}.

The values of the optical opacity of the ejecta $\kappa$ and the $\gamma$-ray opacity $\kappa_{\gamma}$
are set to be 0.07 cm$^2$g$^{-1}$ and 0.027 cm$^2$g$^{-1}$, respectively.
The free parameters of the \Ni model adopted here are the ejecta mass M$_{ej}$,
the ejecta velocity $v_{\rm ph}$,
\footnote{The photospheric velocity of SN~2012au, SN~2013ak, and SN~2013df determined by the
spectral lines are set to be the upper limits of their $v_{\rm ph}$, see Table \ref{Tab}.}
the \Ni mass M$_{\rm Ni}$,
the temperature floor of the photosphere T$_{f}$, the explosion time relative
to the first data t$_{\rm shift}$. \footnote{The date of the first photometric data of every SN is set to be
the day 0.} The additional parameters required
by the cooling plus \Ni model are the extended material
mass M$_{\rm e}$, the extended material radius R$_{\rm e}$, the energy imparted
by the shock passing through the extended material E$_{\rm e}$.

The Markov Chain Monte Carlo (MCMC) using the \texttt{emcee} Python package \citep{Foreman-Mackey2013}
is adopted to get the best-fitting parameters. We employ 20 walkers, each walker runs 30,000 steps.
The uncertainties are 1$\sigma$ confidence, corresponding to the 16th and 84th percentiles of the posterior samples.

\begin{figure}
	\centering
	\includegraphics[width=0.32\textwidth,angle=0]{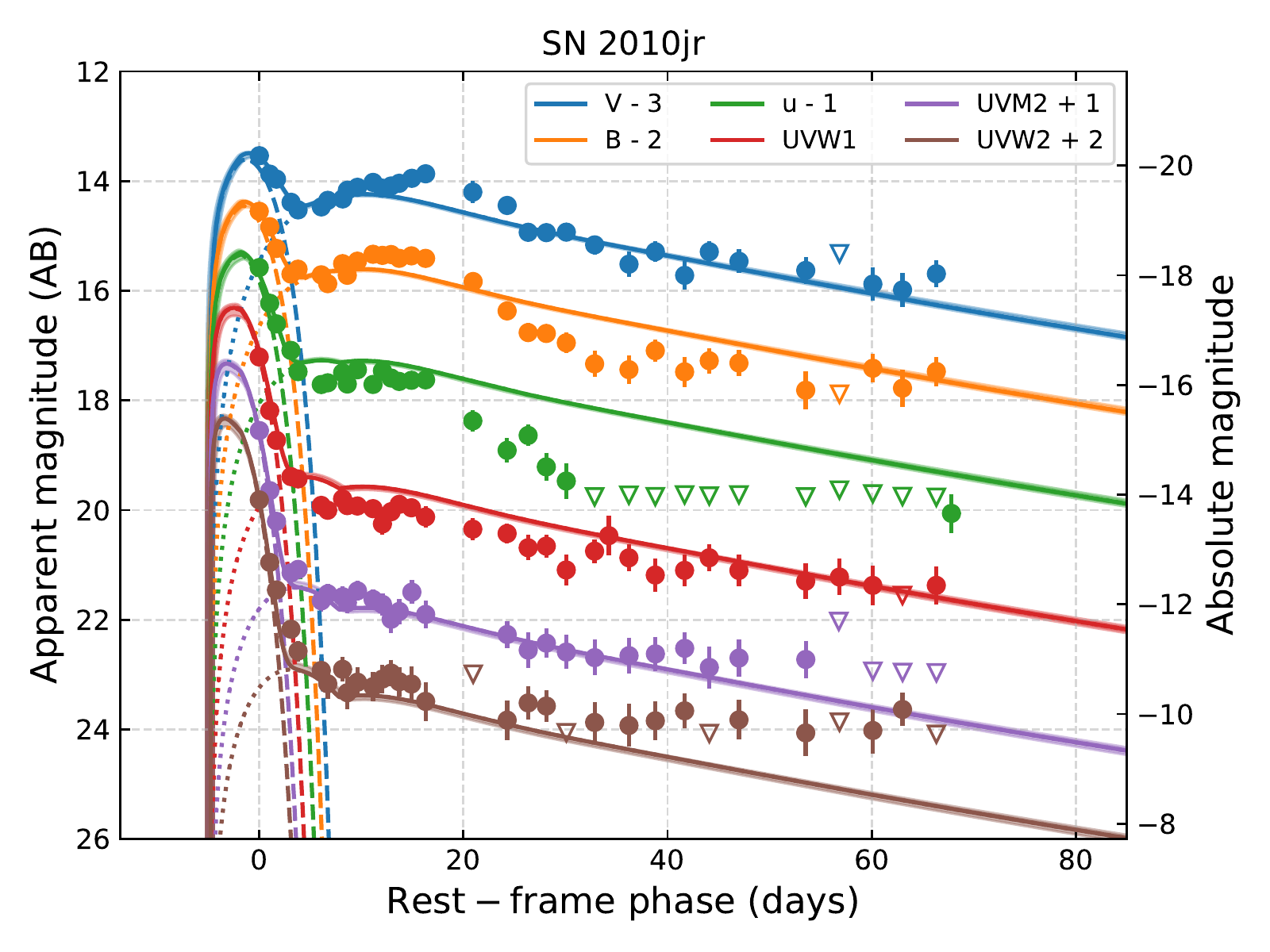}
	\includegraphics[width=0.32\textwidth,angle=0]{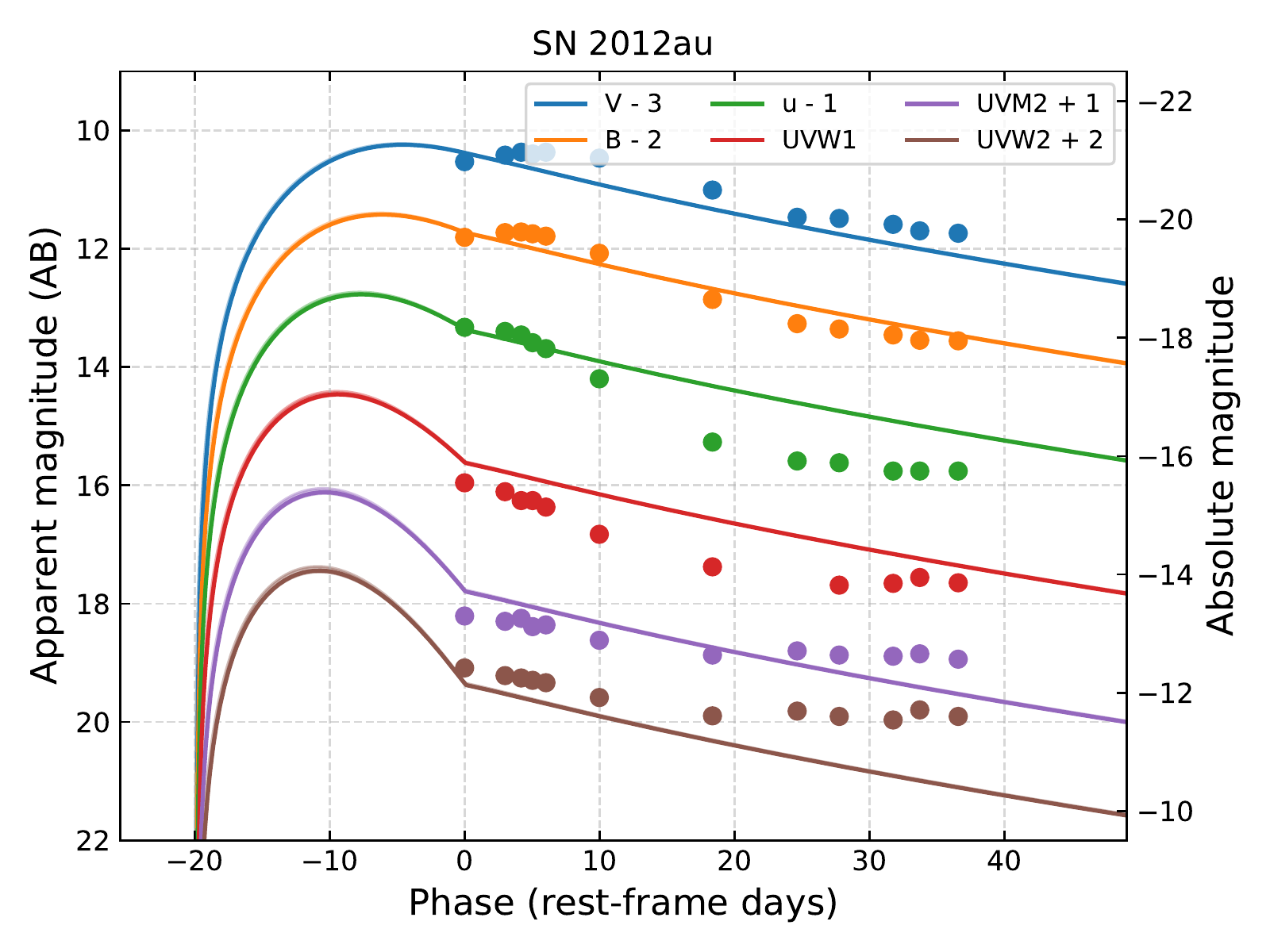}
	\includegraphics[width=0.32\textwidth,angle=0]{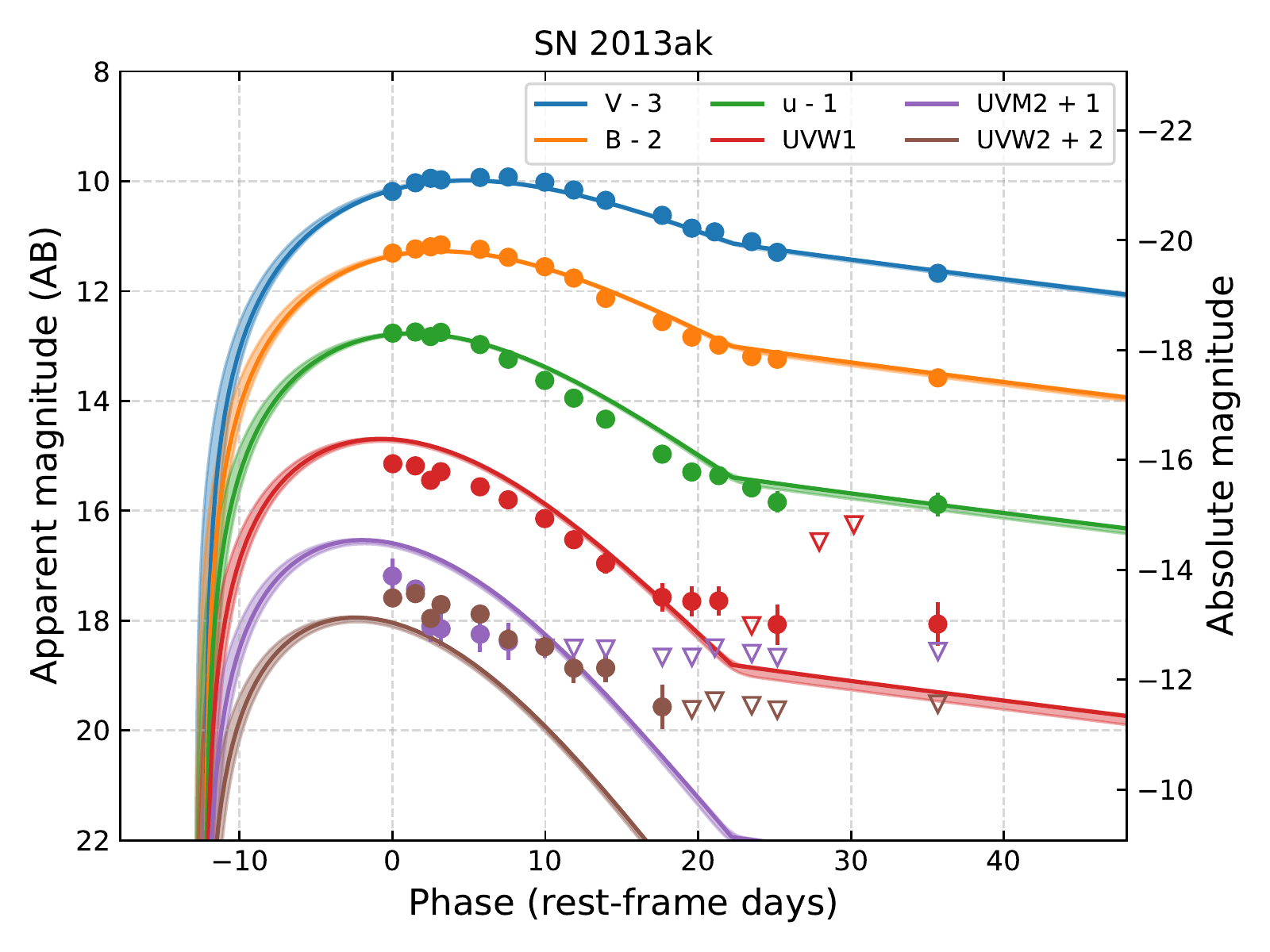}
	\includegraphics[width=0.32\textwidth,angle=0]{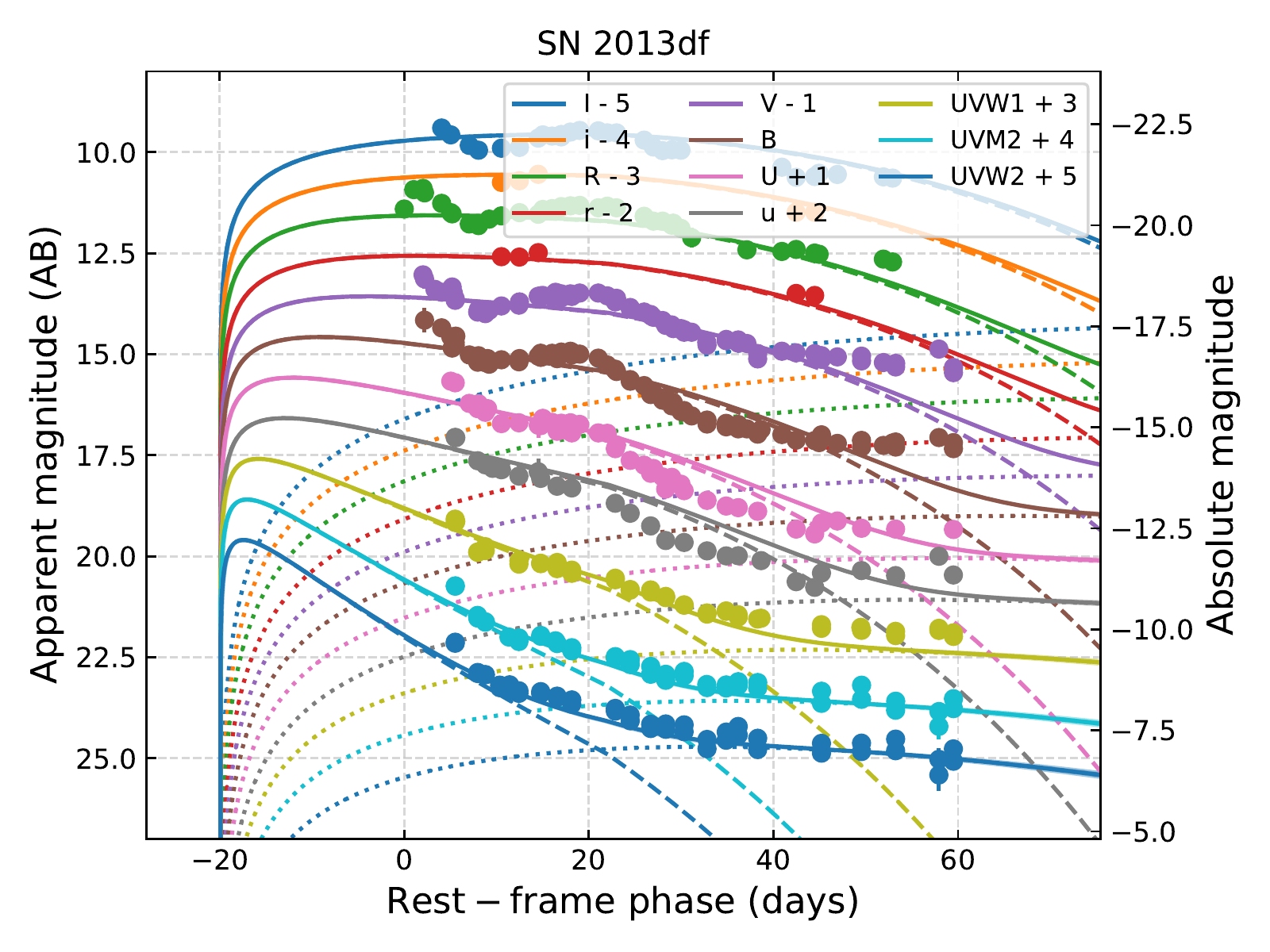}
	\includegraphics[width=0.32\textwidth,angle=0]{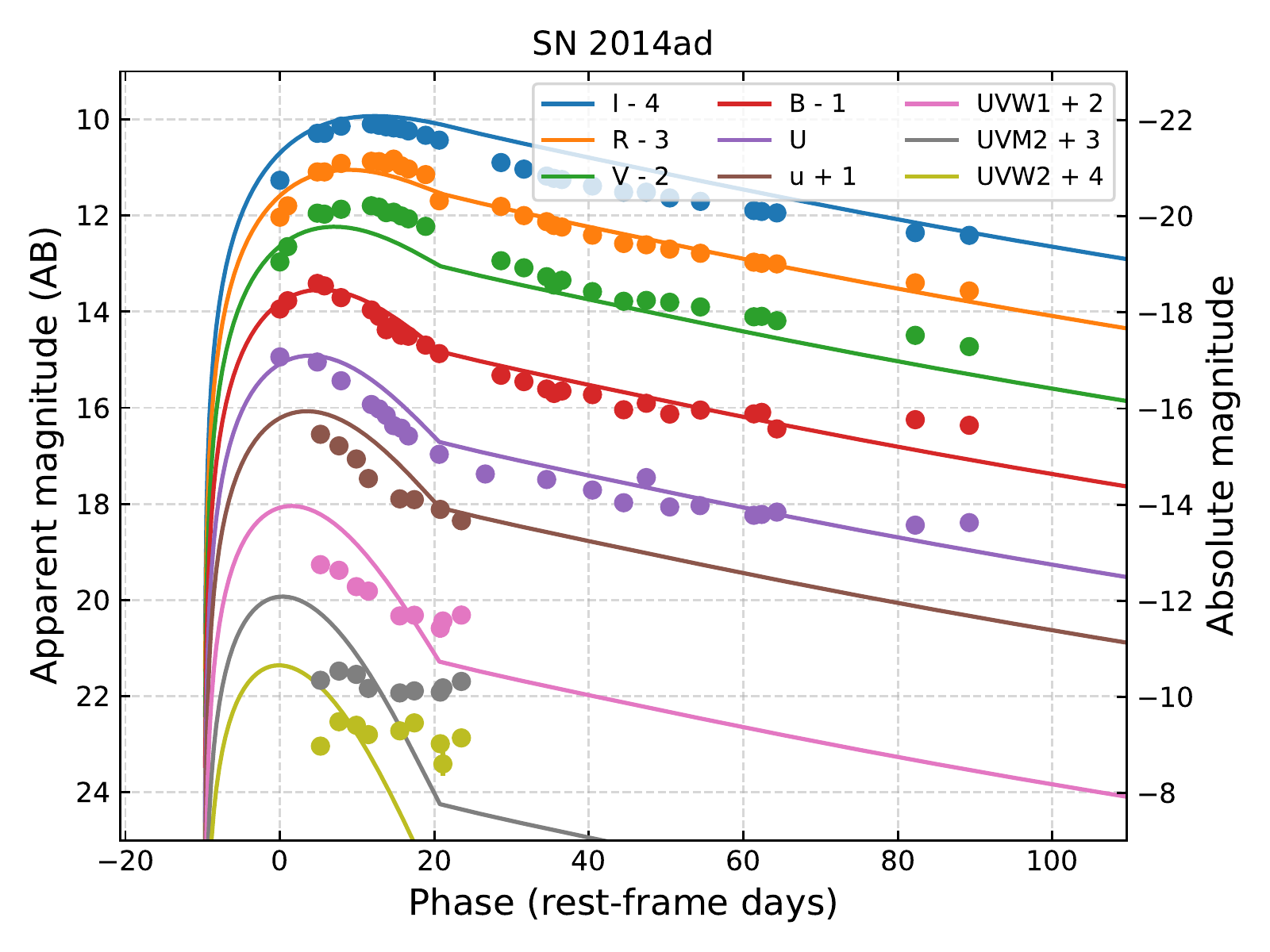}
	\caption{The best fits of the multi-band LCs of SNe (solid lines). The \Ni model is adopted for SN~2012au, SN~2013ak, and SN~2014ad, while the
cooling plus \Ni model is adopted for SN~2010jr and SN~2013df. The shaded regions represent the 1$\sigma$ ranges of the parameters.
The dotted lines and dashed lines are the LCs powered by the \Ni and the cooling, respectively. Triangles represent upper limits.}
\label{Fig:allbands}
\end{figure}

The multi-band LC fits of the five SNe are shown in Figure \ref{Fig:allbands},
and the corresponding optimal parameters are shown in Table \ref{Tab}.
We find that the LCs of SN~2010jr and SN~2012au at all bands cannot be fitted,
the LCs of SN~2013df at optical bands and $U/u$ cannot be fitted,
the optical LCs of SN~2013ak and SN~2014ad can be fitted while their
UV LCs show UV excesses in $UVW1$ and $UVW2$ or $UVW1$, $UVM2$, and $UVW2$ bands at some epochs.

\begin{figure}
	\centering
	\includegraphics[width=0.32\textwidth,angle=0]{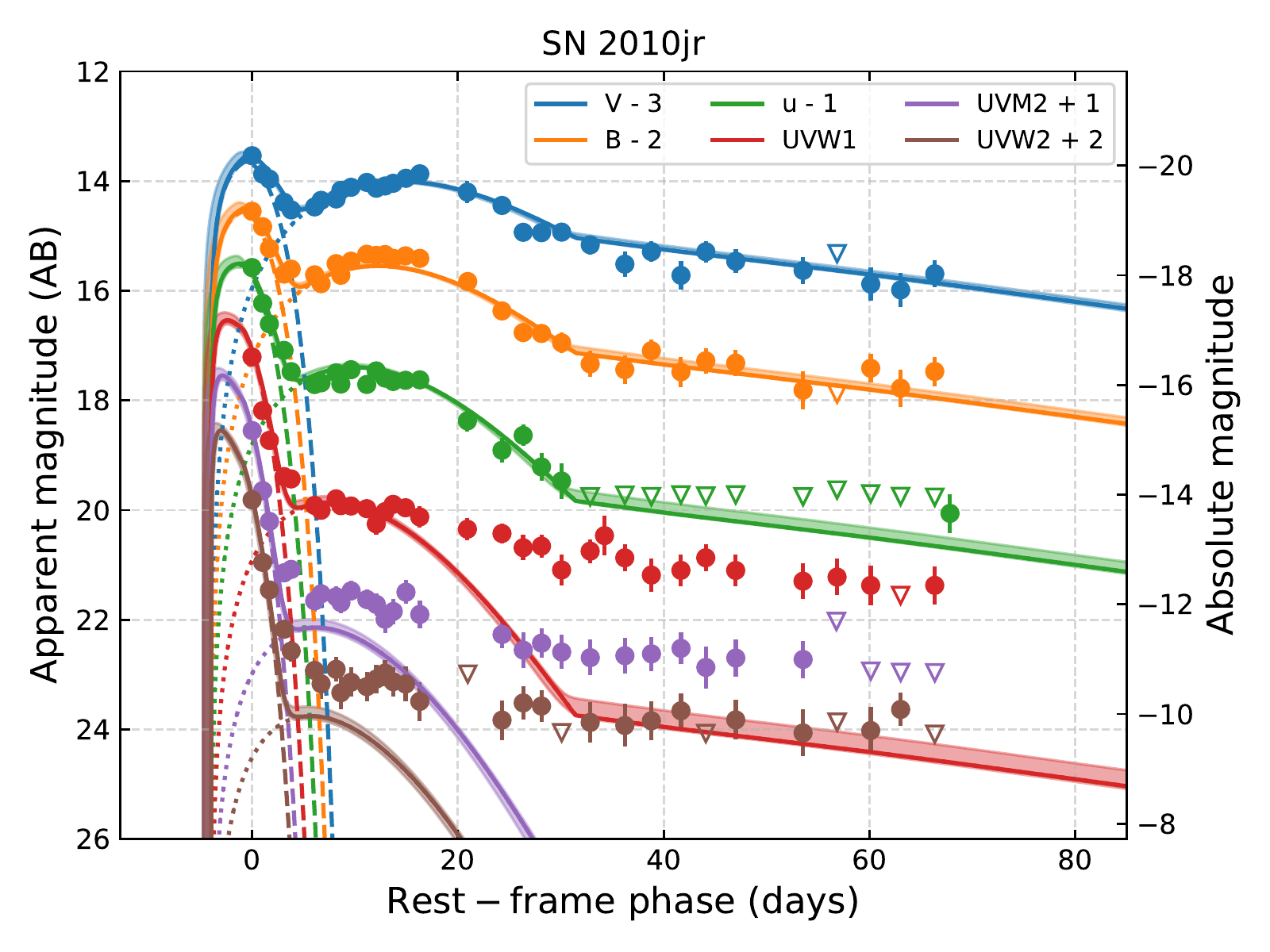}
	\includegraphics[width=0.32\textwidth,angle=0]{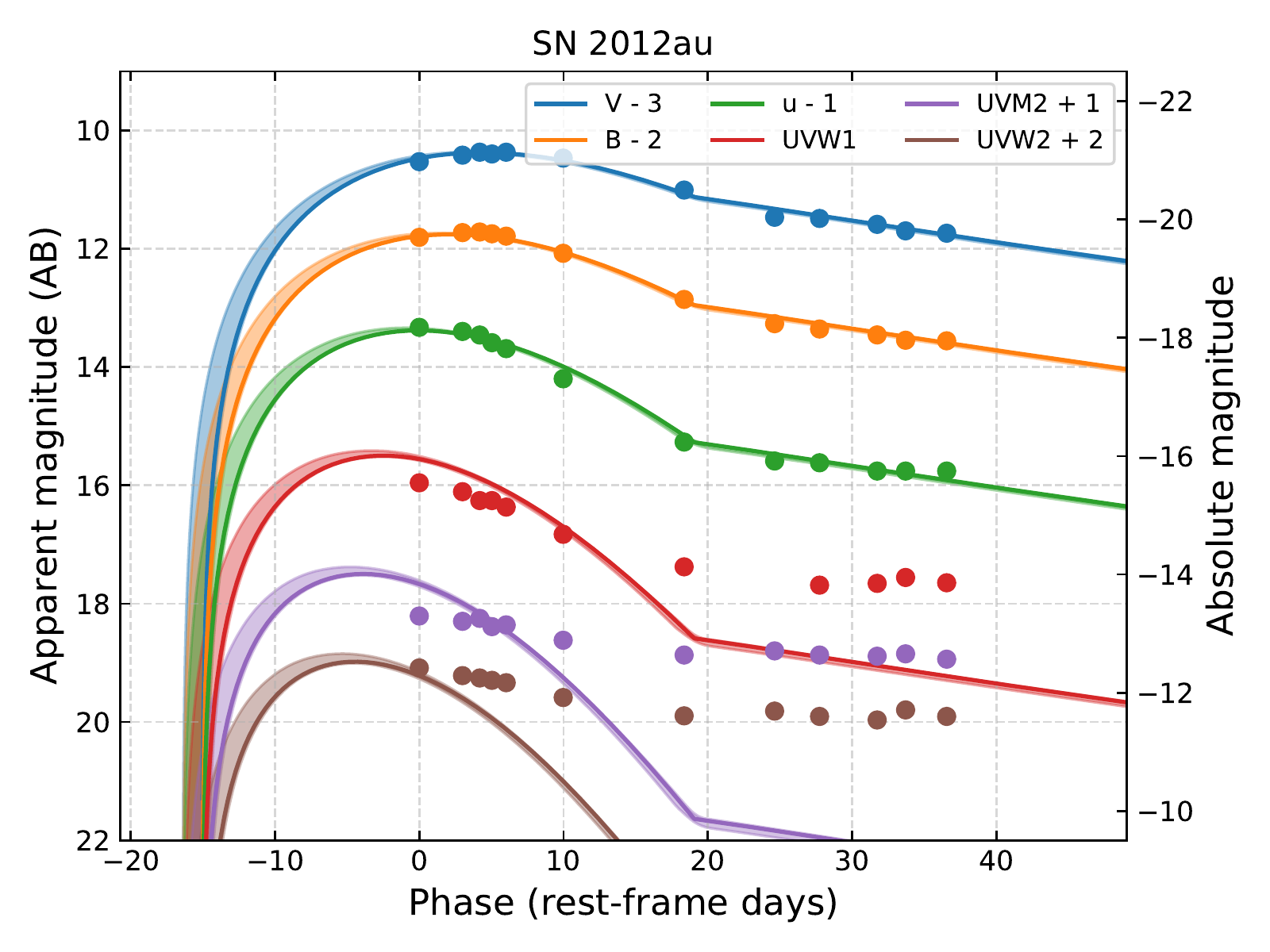}
	\includegraphics[width=0.32\textwidth,angle=0]{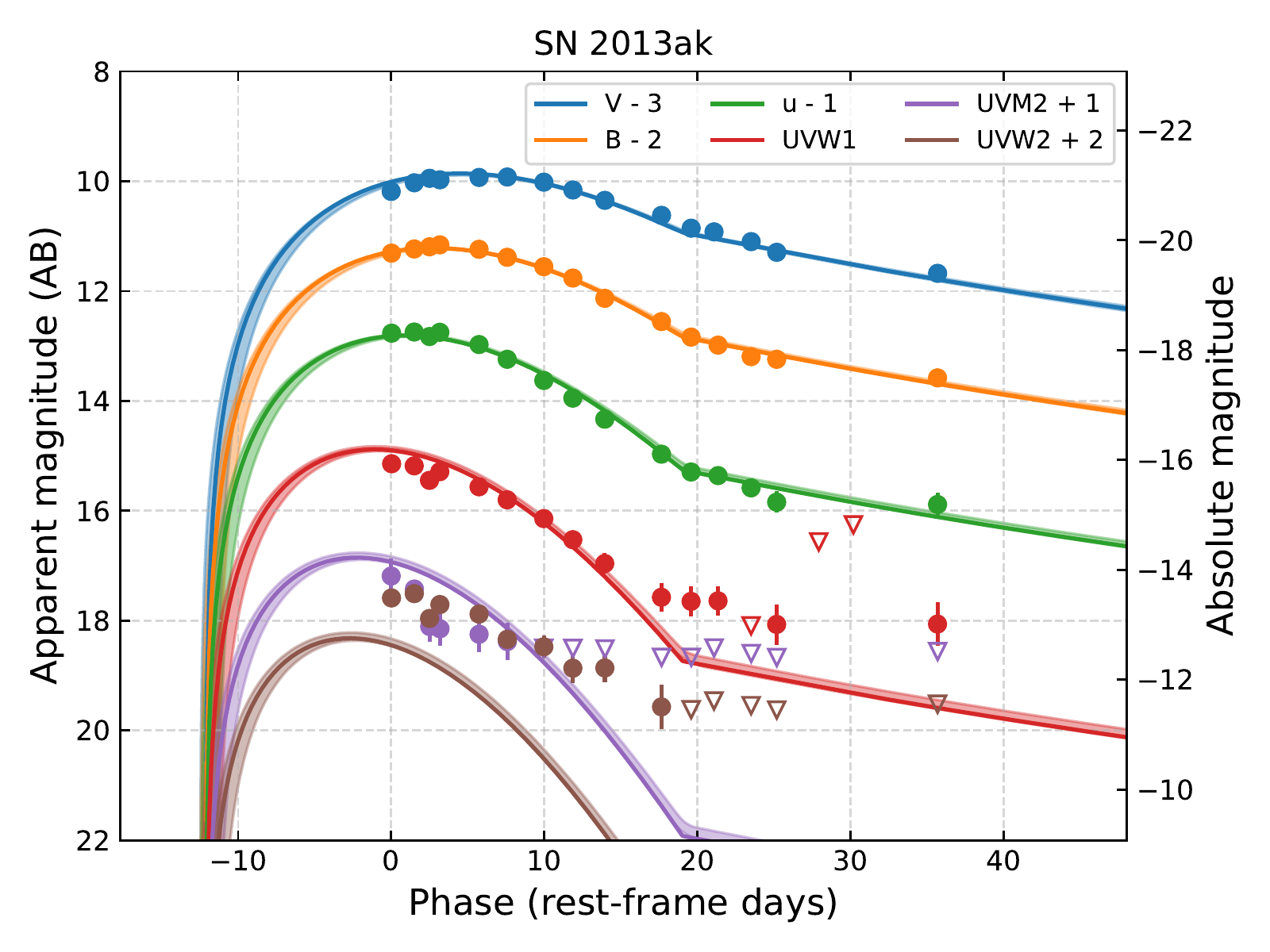}
	\includegraphics[width=0.32\textwidth,angle=0]{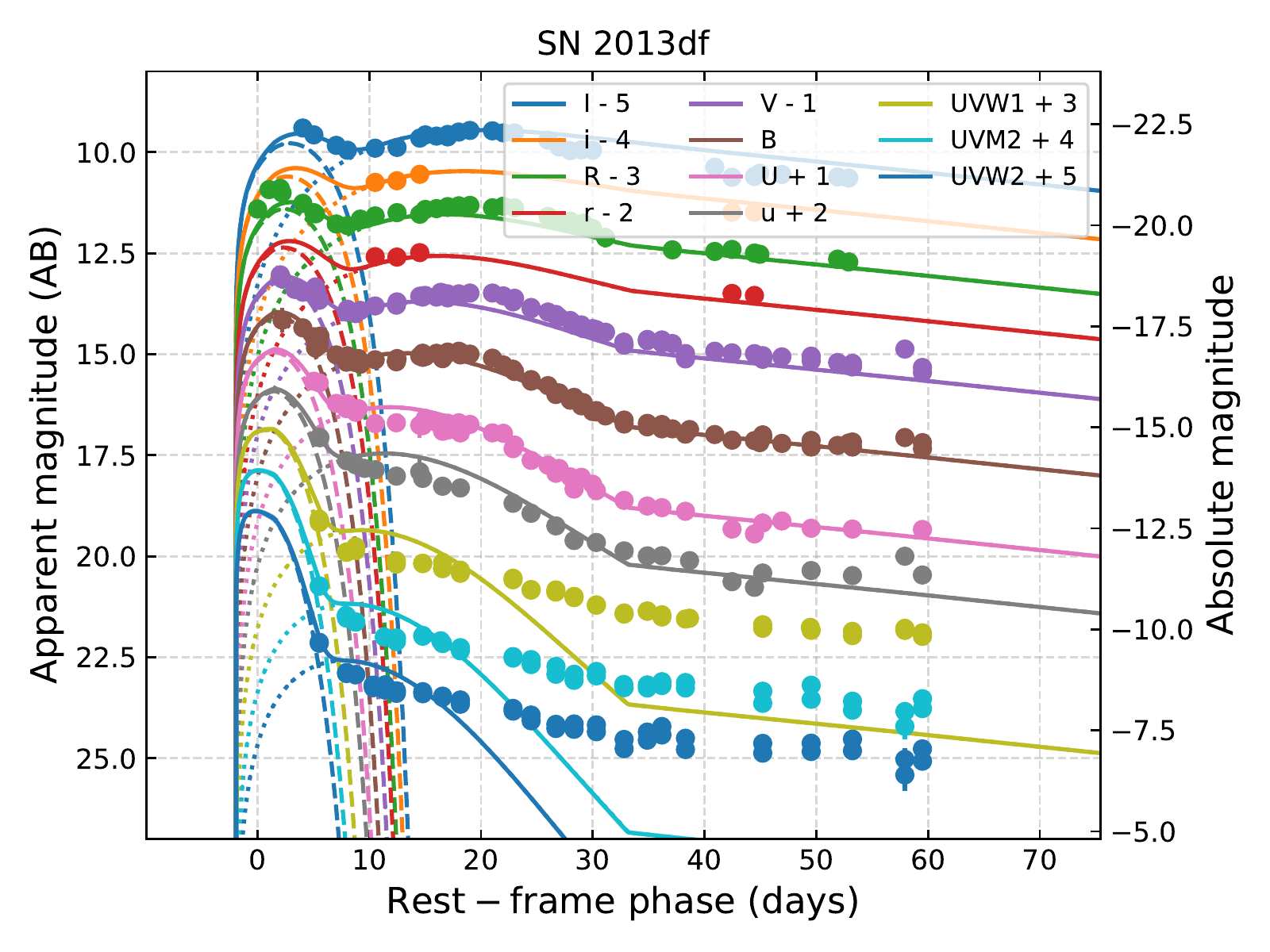}
	\includegraphics[width=0.32\textwidth,angle=0]{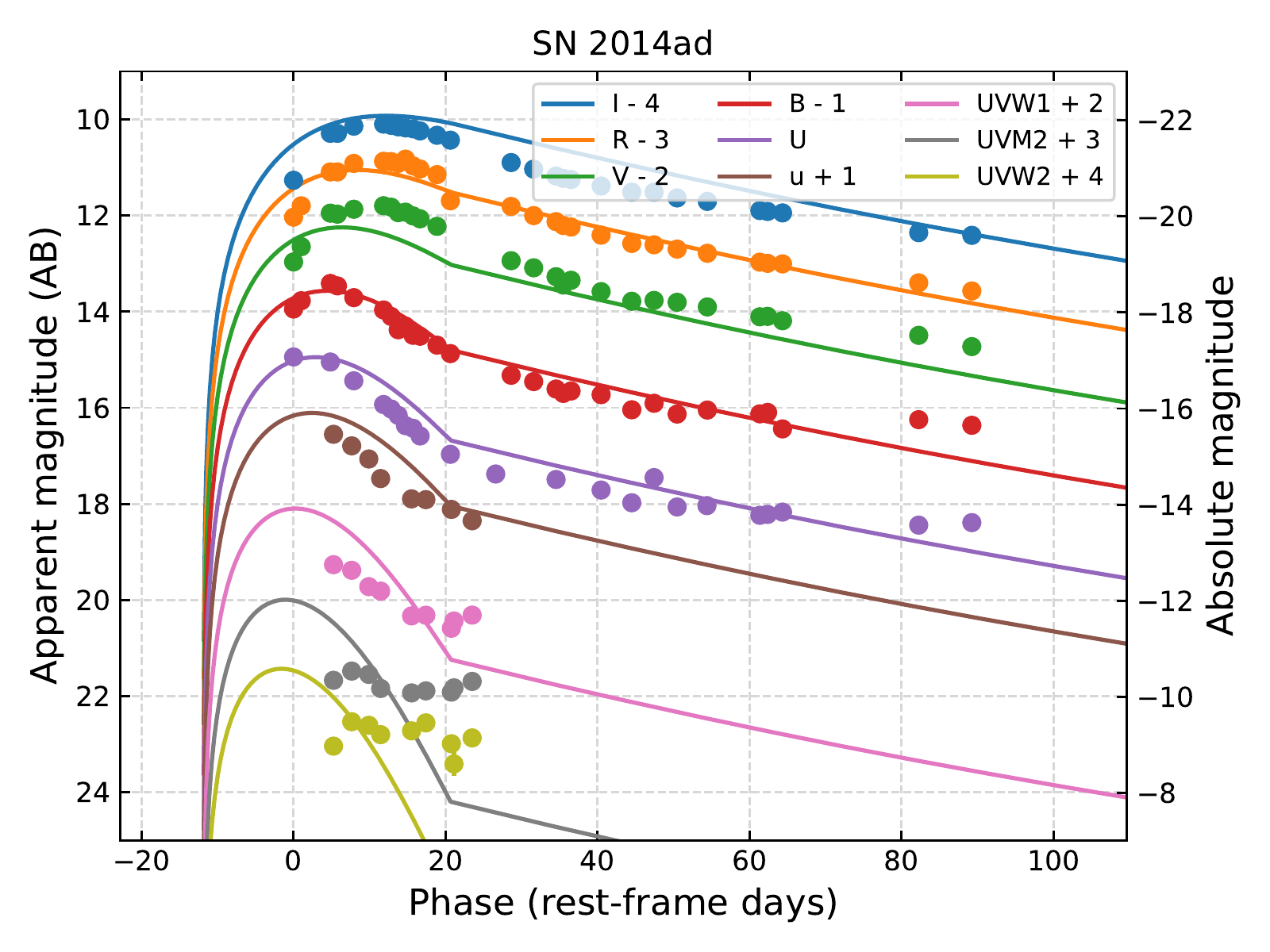}
	\caption{Same as Figure \ref{Fig:allbands}, but $UVW1$, $UVM2$, and $UVW2$ bands are excluded in the fits.}
\label{Fig:opticalbands}
	
\end{figure}

Alternatively, we use the same models to fit the optical and $U/u$ LCs of the five
SNe, see Figure \ref{Fig:opticalbands} (the $UVW1$-, $UVM2$-, and $UVW2$-band photometry
are also plotted, but not fitted) and Table \ref{Tab} for the fits and
the best-fit parameters, respectively. We find that the optical and $U/u$ LCs of
SN~2010jr, SN~2012au, and SN~2013df can be well fitted, while the $UVW1$-, $UVM2$-, and $UVW2$-bands
show apparent UV excesses at most epochs.
Compared to the fits for the data sets in all bands, the fits for SN~2013ak and SN~2014ad do not change
significantly, because the weight of the data in $UVW1$-, $UVM2$-, and $UVW2$-bands is significantly lower than
that of the optical and $U/u$ bands.

The fits for two different data sets (all bands versus optical and $U/u$ bands) also indicate that the SEDs
of the all five SNe show UV excesses, and that the LCs cannot be
fitted by the multi-band models (since the $UVW1$-, $UVM2$-, and $UVW2$-band data cannot be fitted).

\section{Constructing and Fitting the bolometric LCs}
\label{bolometric_LCs}

\begin{figure}
	\centering
	\includegraphics[width=0.32\textwidth,angle=0]{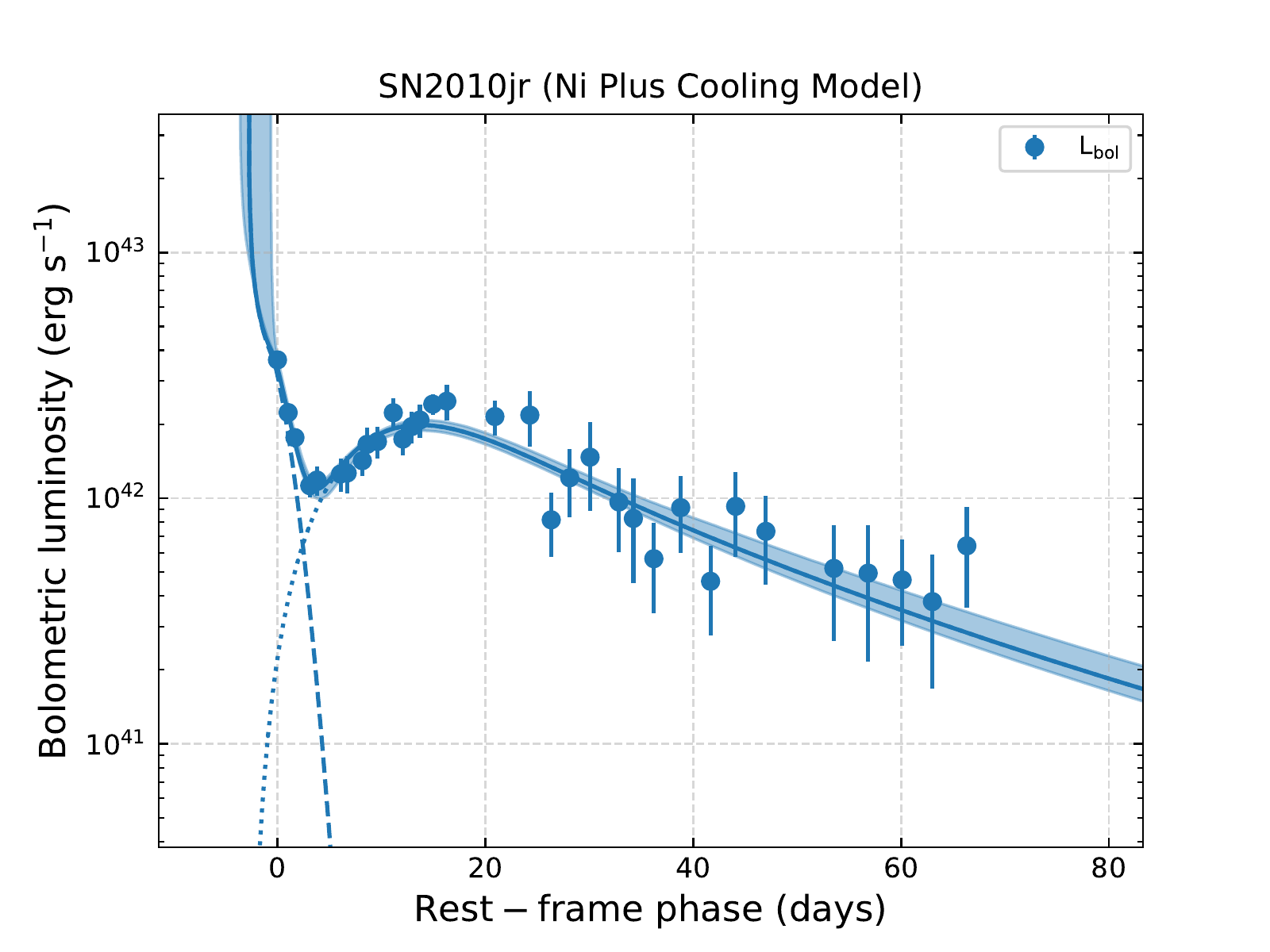}
	\includegraphics[width=0.32\textwidth,angle=0]{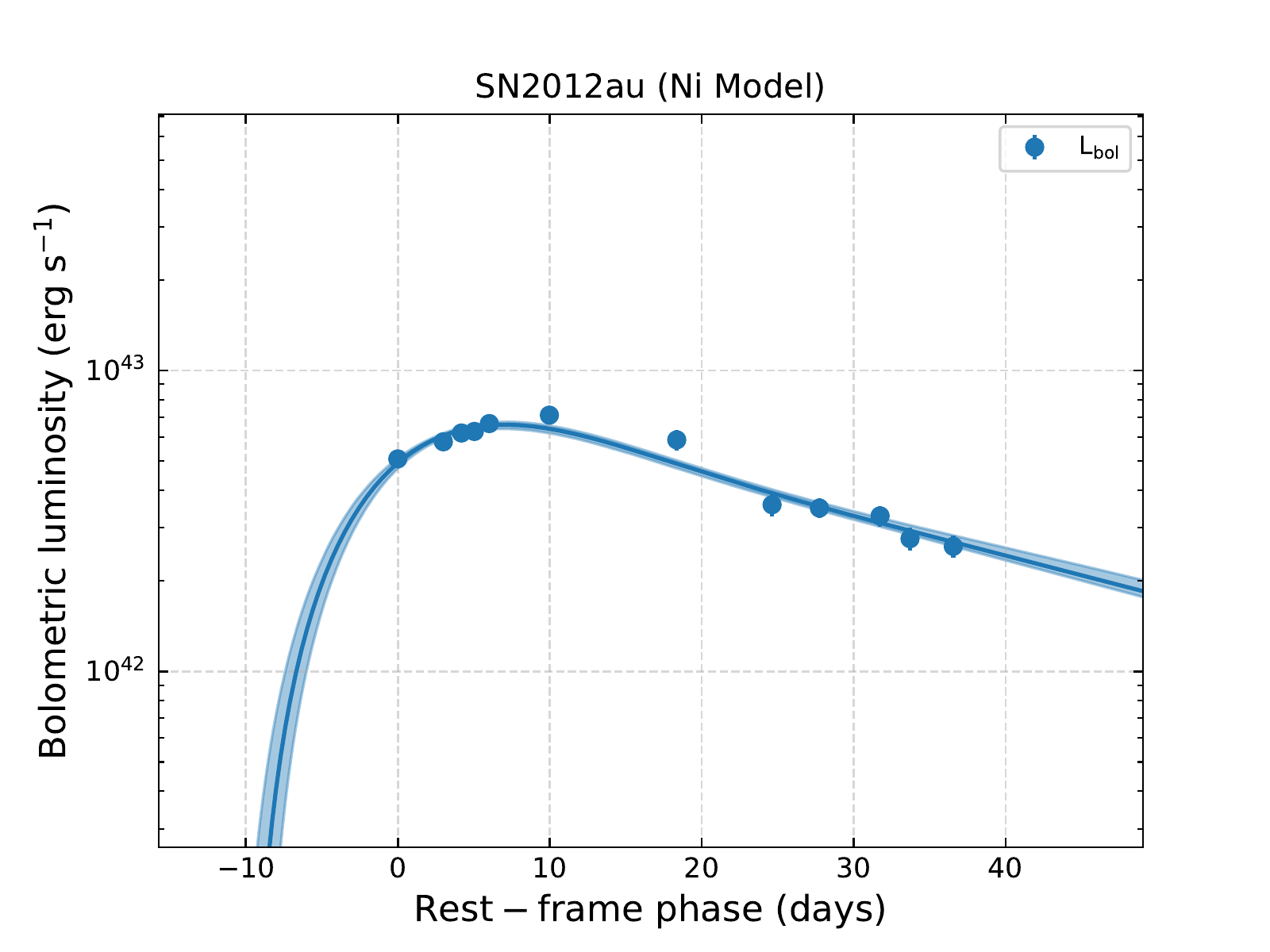}
	\includegraphics[width=0.32\textwidth,angle=0]{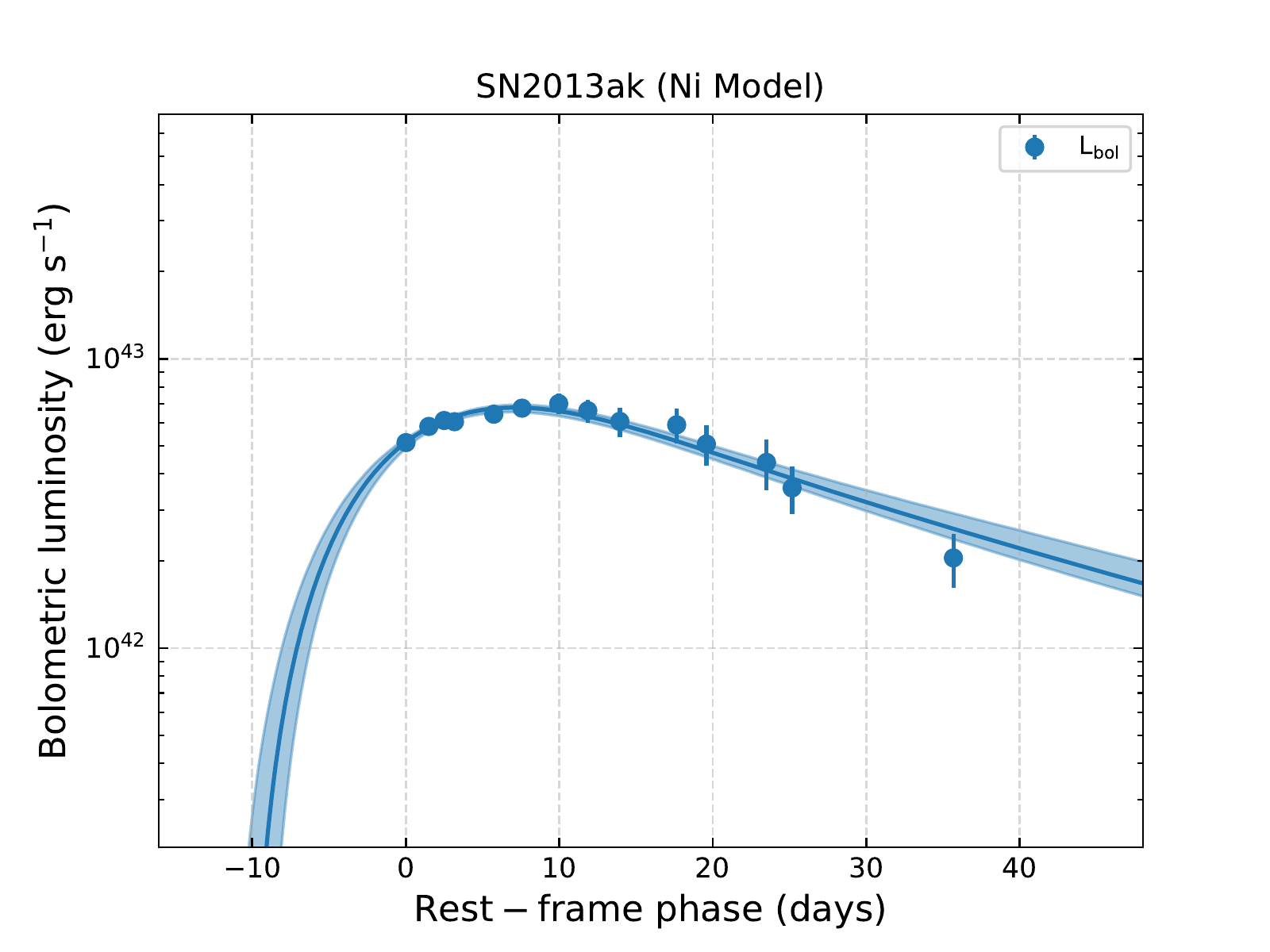}	
	\includegraphics[width=0.32\textwidth,angle=0]{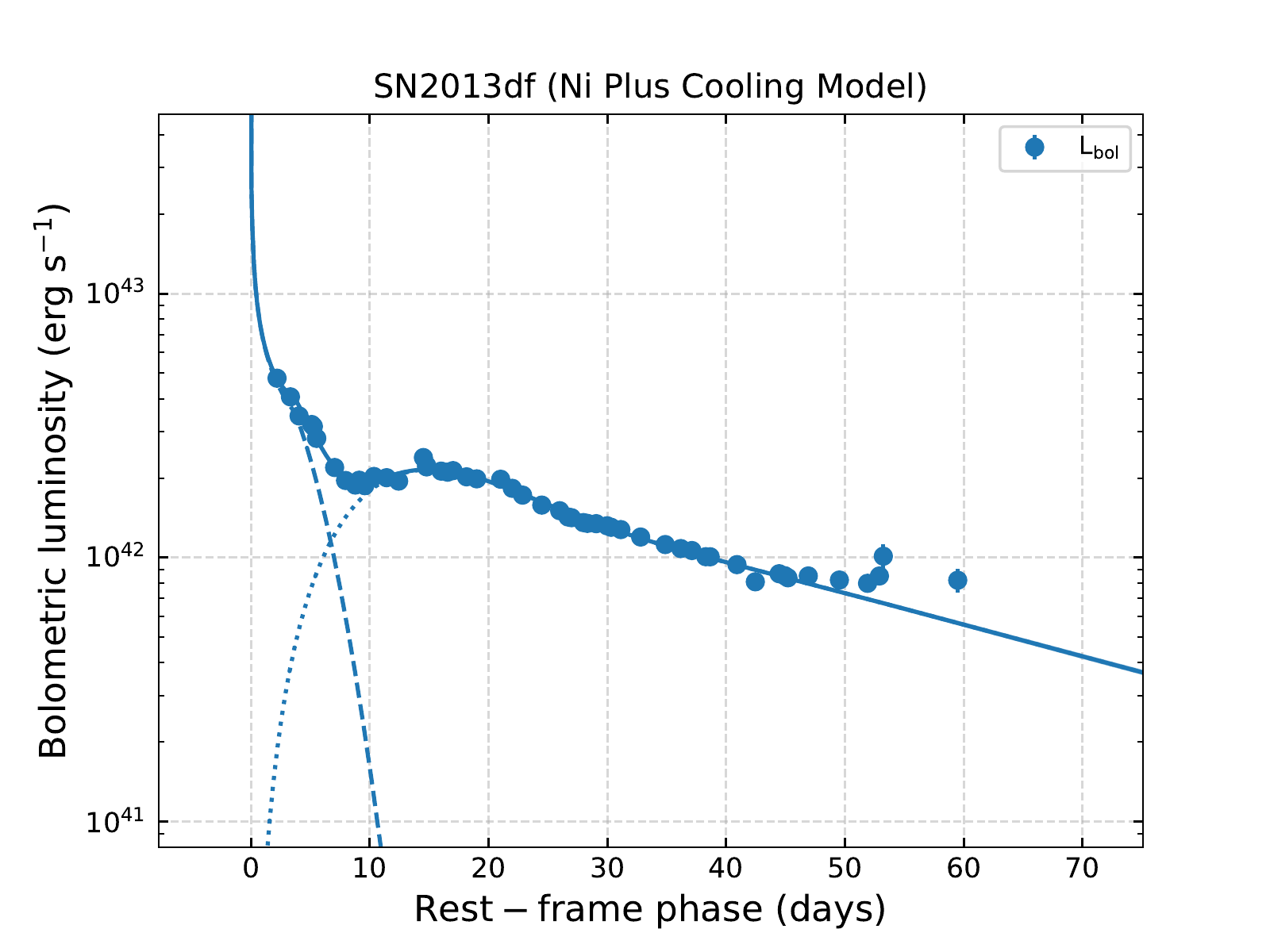}
	\includegraphics[width=0.32\textwidth,angle=0]{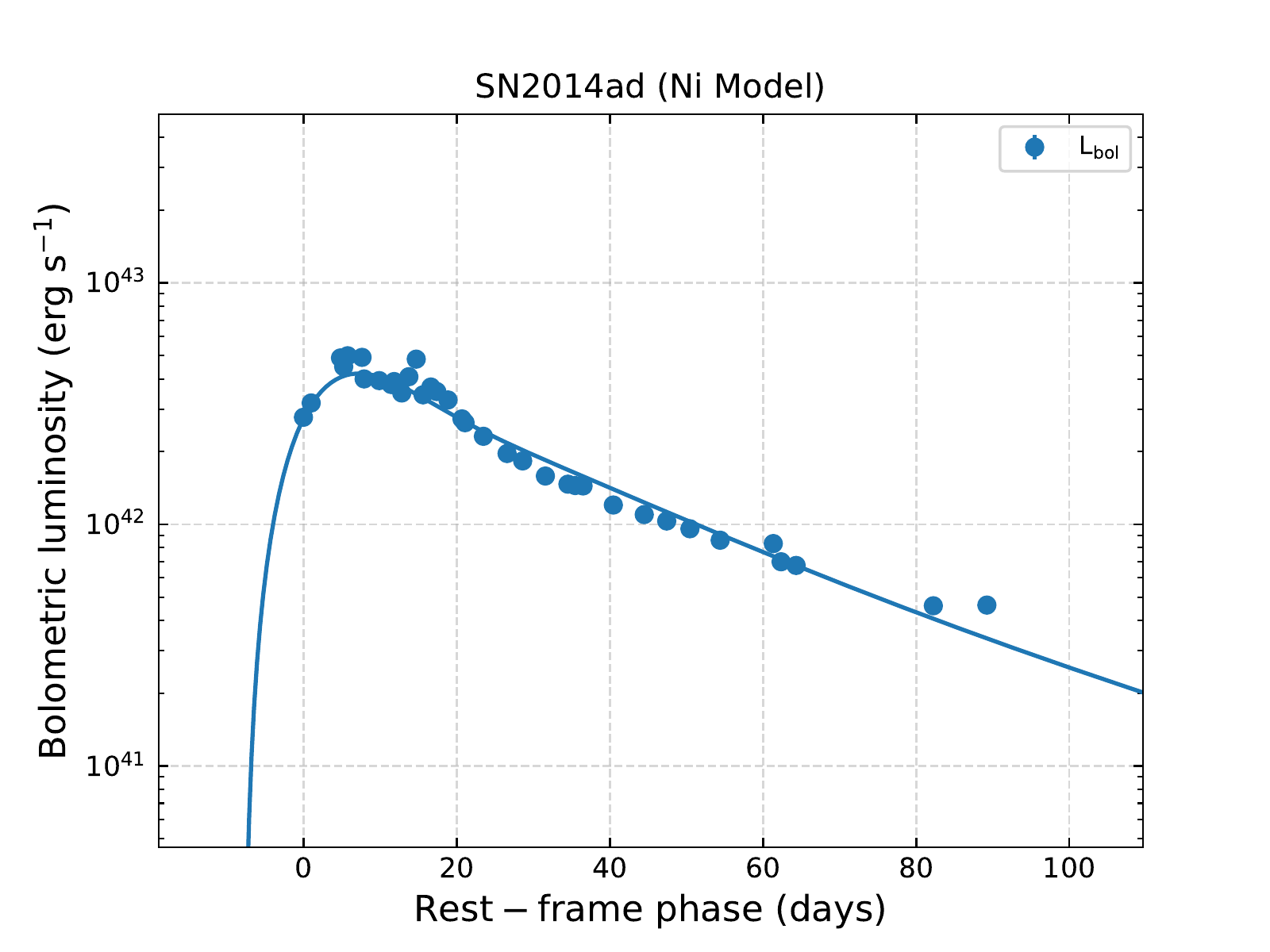}
	\caption{The best fits of the bolometric LCs of SNe (solid lines). The \Ni model is adopted for SN~2012au, SN~2013ak, and SN~2014ad, while the
cooling plus \Ni model is adopted for SN~2010jr and SN~2013df. The shaded regions represent the 1$\sigma$ ranges of the parameters.
The dotted lines and dashed lines are the LCs powered by the \Ni and the cooling, respectively. Triangles represent upper limits.}
\label{Fig:Lbot}
\end{figure}

To constrain the physical properties of the SNe, we must synthesize and fit
the bolometric LCs. To avoid missing the UV and
IR flux, we first divide the area under a theoretical SED into three classes:
(1) The part between 0 to the $\lambda_{\rm UVW2}$ is assumed to be a triangle;
(2) the part between the $\lambda_{\rm UVW2}$ to $\lambda_{\rm u}$ are assumed to
be several trapezoids; (3) the part of a SED longer than $\lambda_{\rm u}$ is described by
the blackbody fit ($f_{\nu,\lambda>\lambda_{\rm u}}=(2\pi h\nu^3/c^2)(e^{{h\nu}/{k_bT_{\rm ph}}}-1)^{-1}\frac{R_{\rm ph}^2}{D^2_{\rm L}}$).
We integrate the three parts of the SEDs and get the flux at all epochs, i.e.,
$f=\int_0^{\infty}f_{\nu}d{\nu}=\int_0^{\lambda_{\rm UVW2}}f_{\nu} d\nu + \int_{\lambda_{\rm UVW2}}^{\lambda_{\rm u}}f_{\nu} d\nu + \int_{\lambda_{\rm u}}^{\infty}f_{\nu}d\nu$.
The bolometric luminosities can be calculated by $L_{\rm bol}=4{\pi}D^2_{\rm L}f$.
\footnote{The SEDs of SN~2014ad after day 23 do not have
$u$, $UVW1$, $UVM2$, and $UVW2$ photometry. We roughly assume that the
ratio of the luminosity of the part between 0 to $\lambda_{\rm u}$
to bolometric luminosity is constant ($\sim$6\% which is the mean value derived from the
integration for the early-epoch SEDs).}
The synthesized bolometric LCs of the five SNe are shown in Figure \ref{Fig:Lbot}.

The parameters of \Ni model used to fit the bolometric LCs of SN~2012au, SN~2013ak, and SN~2014ad
and the cooling plus \Ni model used to fit the bolometric LCs of SN~2010jr and SN~2013df are listed in
Table \ref{Tab}. The bolometric LC fits and the best-fit parameters are shown in Figure \ref{Fig:Lbot} and
Table \ref{Tab}, respectively. We find that the bolometric LCs can be well fitted by the
models and the values of $\chi^{2}/$dof (dof=degree of freedom) are significantly smaller than
the multi-band LC fits (see the last column of Table \ref{Tab}).
This indicates that the method of synthesizing and fitting the bolometric LCs is the best one
for the SNe showing UV excesses.

It should be noted that, however, the synthesized bolometric LC of SN~2014ad around the peak
show undulation feature (see Figure \ref{Fig:Lbot}). This is due to the undulation of the photometric data in
$UVW1$, $UVM2$, and $UVW2$ bands (see Figures \ref{Fig:allbands} and \ref{Fig:opticalbands}).
The undulation results in a large $\chi^{2}/$dof. Nevertheless, the value of $\chi^{2}/$dof of
the fit of the bolometric LC is significantly smaller than that of the fit of the multi-band LCs.

\section{Discussion and Conclusions}
\label{sec:discussion}

In this paper, we demonstrate that the SEDs of five SNe show UV excesses by
fitting their SEDs and the multi-band LCs. We suggest that the five SNe and
other optical transients having UV excesses cannot be fitted by the multi-band
models including a UV-absorbed or standard blackbody module.

To constrain the physical properties of the SNe showing UV excesses, we
construct and fit their bolometric LCs, getting decent fits.
Moreover, the bolometric LC fits are significantly better than the
multi-band fits. Our results indicate that the most reasonable scheme of deriving the physical
properties of the optical transients having UV excesses might be constructing
and fitting the bolometric LCs, because the multi-band LCs cannot be fitted.

It is important to emphasize that the multi-band models including a (UV-absorbed)
blackbody module are still the best ones to fit the optical transients which
show UV absorption or have very few or no UV data. In fact, the reliable bolometric
LCs are difficult to be constructed for the optical transients with sparse or no UV data.

Currently, only a small fraction of SNe, TDEs, and other optical transients have been
observed in UV bands shorter than 3000 \AA. In the future, the Large Synoptic Survey Telescope (LSST)
and other large survey telescope would obtain the far UV photometry of numerous optical
transients at high redshifts (e.g., $z\gtrsim1$) using the optical filters, since the flux
in some UV bands would be redshifted and become optical flux observed.
Therefore, we can expect that the large optical survey telescopes
would discover a great number of optical transients showing UV excesses, which cannot
be fitted by the multi-band models that include (UV-absorbed) blackbody module,
but can be well modeled by constructing and fitting their bolometric LCs.

\acknowledgments

This work is supported by National Natural Science Foundation of China (grant 11963001).

\clearpage

\begin{center}
		\setlength{\tabcolsep}{0.05mm}{
		
			\begin{longtable}{ccccccccccc}
				\caption{The best fitting parameters of the multi-band and bolometric LC models.}\\
				\label{Tab} \\
				\hline
				\toprule
				\hline
				Name&~~M$_{\rm e}$&~~R$_{e}$&lgE$_{e,50}$&~~M$_{\rm ej}$&~~$v_{\rm ph}$&~~M$_{\rm Ni}$&~~T$_{\rm f}$&~~t$_{\rm shift}$&~~$\chi^{2}/$dof~~\\
				unit&~~(M$_{\odot}$)&~~(R$_{\odot}$)&~~($10^{50} \rm erg$)&~~(M$_{\odot}$)&~~($10^9\rm cm\ s^{-1}$)&~~(M$_{\odot}$)&~~(K)&~~(day)&~~\\
				\hline
				Prior&~~[0.01,20]&~~[10,3000]&~~[-3,3]&~~[0.1,50]&~~[0.1,A$^a$]&~~[0.001,2]&~~[1000,10000]&~~[-20,0]&~~\\	
				\hline
				\multicolumn{11}{c}{parameters of multi-band LC fits}\\
				\hline
				\toprule
				SN~2010jr&~~0.04$^ {+0.00 }_ {-0.00}$&~~121.49$^ {+45.63 }_ {-35.60}$&~~-0.67$^ {+0.12 }_ {-0.11}$&~~1.58$^ {+0.06 }_ {-0.06}$&~~0.91$^ {+0.03 }_ {-0.03}$&~~0.05$^ {+0.00 }_ {-0.00}$&~~6406.92$^ {+44.73 }_ {-43.18}$&~~-4.87$^ {+0.39 }_ {-0.37}$&~~3.99 ~~\\
				SN~2012au&~~-&~~-&~~-&~~1.84$^ {+0.04 }_ {-0.04}$&~~1.25$^ {+0.00 }_ {-0.01}$&~~0.27$^ {+0.00 }_ {-0.00}$&~~6524.76$^ {+12.97 }_ {-13.28}$&~~-19.89$^ {+0.10 }_ {-0.05}$&~~60.52 ~~\\
				SN~2013ak&~~-&~~-&~~-&~~3.38$^ {+0.11 }_ {-0.11}$&~~1.33$^ {+0.04 }_ {-0.04}$&~~0.27$^ {+0.00 }_ {-0.00}$&~~4168.13$^ {+45.77 }_ {-42.59}$&~~-12.39$^ {+0.49 }_ {-0.49}$&~~10.46 ~~\\
				SN~2013df&~~1.97$^ {+0.01 }_ {-0.01}$&~~439.65$^ {+4.06 }_ {-4.16}$&~~0.01$^ {+0.01 }_ {-0.01}$&~~0.33$^ {+0.04 }_ {-0.03}$&~~0.01$^ {+0.00 }_ {-0.00}$&~~0.01$^ {+0.00 }_ {-0.00}$&~~5491.97$^ {+3050.79 }_ {-3044.92}$&~~-19.93$^ {+0.10 }_ {-0.05}$&~~67.68 ~~\\
				SN~2014ad&~~-&~~-&~~-&~~3.05$^ {+0.01 }_ {-0.01}$&~~1.23$^ {+0.00 }_ {-0.00}$&~~0.20$^ {+0.00 }_ {-0.00}$&~~4498.80$^ {+2.71 }_ {-2.69}$&~~-9.77$^ {+0.04 }_ {-0.04}$&~~650.11 ~~\\
				\hline
				\hline
				\multicolumn{11}{c}{parameters of multi-band LC fits (excluding $UVW1$, $UVM2$, and $UVW2$ bands}\\
				\hline
				\toprule
				SN~2010jr&~~0.05$^ {+0.01 }_ {-0.00}$&~~63.41$^ {+22.95 }_ {-17.90}$&~~-0.39$^ {+0.13 }_ {-0.10}$&~~5.24$^ {+0.43 }_ {-0.40}$&~~1.13$^ {+0.07 }_ {-0.07}$&~~0.12$^ {+0.01 }_ {-0.01}$&~~3738.79$^ {+71.25 }_ {-67.01}$&~~-4.33$^ {+0.43 }_ {-0.46}$&~~42.92 ~~\\
				SN~2012au&~~-&~~-&~~-&~~3.72$^ {+0.13 }_ {-0.13}$&~~1.24$^ {+0.01 }_ {-0.02}$&~~0.35$^ {+0.01 }_ {-0.01}$&~~4341.63$^ {+29.95 }_ {-29.60}$&~~-18.04$^ {+0.49 }_ {-0.53}$&~~460.30 ~~\\
				SN~2013ak&~~-&~~-&~~-&~~3.01$^ {+0.16 }_ {-0.16}$&~~1.65$^ {+0.07 }_ {-0.07}$&~~0.31$^ {+0.01 }_ {-0.01}$&~~4165.86$^ {+47.82 }_ {-49.10}$&~~-11.92$^ {+0.51 }_ {-0.57}$&~~12.68 ~~\\
				SN~2013df&~~0.05$^ {+0.00 }_ {-0.00}$&~~167.26$^ {+18.84 }_ {-16.58}$&~~-0.41$^ {+0.04 }_ {-0.04}$&~~1.95$^ {+0.02 }_ {-0.02}$&~~0.84$^ {+0.01 }_ {-0.01}$&~~0.11$^ {+0.00 }_ {-0.00}$&~~4150.27$^ {+7.02 }_ {-6.80}$&~~-1.97$^ {+0.08 }_ {-0.08}$&~~204.13 ~~\\
				SN~2014ad&~~-&~~-&~~-&~~2.50$^ {+0.01 }_ {-0.01}$&~~1.15$^ {+0.00 }_ {-0.00}$&~~0.22$^ {+0.00 }_ {-0.00}$&~~4510.48$^ {+2.88 }_ {-2.92}$&~~-11.88$^ {+0.05 }_ {-0.04}$&~~650.29 ~~\\
				\hline
				\hline
				\multicolumn{11}{c}{parameters of bolometric LC fits}\\
				\hline
				\toprule
				SN~2010jr&~~0.01$^ {+0.01 }_ {-0.00}$&~~478.05$^ {+593.66}_ {-339.60}$&~~-2.08$^ {+0.82 }_ {-0.54}$&~~3.71$^ {+1.06 }_ {-1.23}$&~~1.67$^ {+0.24 }_ {-0.41}$&~~0.09$^ {+0.00 }_ {-0.01}$&~~-&~~-2.78$^ {+2.14 }_ {-0.87}$&~~1.18 ~~\\
				SN~2012au&~~-&~~-&~~-&~~2.70$^ {+0.42 }_ {-0.53}$&~~1.12$^ {+0.10 }_ {-0.18}$&~~0.31$^ {+0.01 }_ {-0.01}$&~~-&~~-10.08$^ {+0.78 }_ {-0.88}$&~~1.72 ~~\\
				SN~2013ak&~~-&~~-&~~-&~~4.20$^ {+1.00 }_ {-1.05}$&~~1.62$^ {+0.17 }_ {-0.30}$&~~0.32$^ {+0.02 }_ {-0.02}$&~~-&~~-10.52$^ {+1.04 }_ {-1.22}$&~~0.53 ~~\\
				SN~2013df&~~0.03$^ {+0.01 }_ {-0.00}$&~~279.51$^ {+116.25 }_ {-91.36}$&~~-1.38$^ {+0.26 }_ {-0.23}$&~~1.26$^ {+0.04 }_ {-0.04}$&~~0.71$^ {+0.02 }_ {-0.02}$&~~0.09$^ {+0.00 }_ {-0.00}$&~~-&~~-0.03$^ {+0.02 }_ {-0.05}$&~~20.74 ~~\\
				SN~2014ad&~~-&~~-&~~-&~~1.66$^ {+0.05 }_ {-0.05}$&~~0.91$^ {+0.02 }_ {-0.02}$&~~0.18$^ {+0.00 }_ {-0.00}$&~~-&~~-7.97$^ {+0.07 }_ {-0.07}$&~~174.86 ~~\\
				\bottomrule
			\end{longtable}}	
		$^a$ \textbf{Note}. For SN~2010jr and SN~2014ad, A$=5$; for SN~2012au, A$=1.25$ \citep{Pandey2021}; for SN~2013ak A$=1.85$ \citep{Carrasco2013}; for SN~2013df, A$=0.9$ \citep{Morales-Garoffolo2014}.
\end{center}

\clearpage

\clearpage

\clearpage

\clearpage

\end{document}